\documentclass[twocolumn,secnumarabic,amssymb, aps, prl, figure, superscriptaddress,floatfix, tightenlines] {revtex4-1}
\usepackage{graphicx}

\begin{document}

\title{Two-dimensional Excitonic Photoluminescence in Graphene on Cu surface}%
\author{Youngsin Park}
\thanks{These authors contributed equally to this work.}
\affiliation{Center for Superfunctional Materials, Department of Chemistry and Department of Physics, Ulsan National Institute of Science and Technology (UNIST), Ulsan 44919, Korea}

\author{Yoo S. Kim} 
\thanks{These authors contributed equally to this work.}
\affiliation{Department of Physics, Sungkyunkwan University, Suwon 16418, Korea}

\author{Chang Woo Myung}
\thanks{These authors contributed equally to this work.} 
\affiliation{Center for Superfunctional Materials, Department of Chemistry and Department of Physics, Ulsan National Institute of Science and Technology (UNIST), Ulsan 44919, Korea}

\author{Robert A. Taylor}
\affiliation{Clarendon Laboratory, Department of Physics, University of Oxford, Parks Road, Oxford OX1 3PU, UK}
\author{Christopher C. S. Chan}
\affiliation{Clarendon Laboratory, Department of Physics, University of Oxford, Parks Road, Oxford OX1 3PU, UK}
\author{Benjamin P. L. Reid}
\affiliation{Clarendon Laboratory, Department of Physics, University of Oxford, Parks Road, Oxford OX1 3PU, UK}
\author{Timothy J. Puchtler}
\affiliation{Clarendon Laboratory, Department of Physics, University of Oxford, Parks Road, Oxford OX1 3PU, UK}
\author{Robin J. Nicholas}
\affiliation{Clarendon Laboratory, Department of Physics, University of Oxford, Parks Road, Oxford OX1 3PU, UK}
\author{Tomba S. Laishram}
\affiliation{Center for Superfunctional Materials, Department of Chemistry and Department of Physics, Ulsan National Institute of Science and Technology (UNIST), Ulsan 44919, Korea}
\author{Geunsik Lee}
\affiliation{Center for Superfunctional Materials, Department of Chemistry and Department of Physics, Ulsan National Institute of Science and Technology (UNIST), Ulsan 44919, Korea}
\author{Chan C. Hwang}
\email{cchwang@postech.ac.kr}
\affiliation{Beamline division, Pohang Accelerator Laboratory, Pohang 37673, Korea}
\author{Chong Yun Park} 
\email{cypark@skku.ac.kr}
\affiliation{Department of Physics, Sungkyunkwan University, Suwon 16418, Korea}
\author{Kwang S. Kim} 
\email{kimks@unist.ac.kr}
\affiliation{Center for Superfunctional Materials, Department of Chemistry and Department of Physics, Ulsan National Institute of Science and Technology (UNIST), Ulsan 44919, Korea}


\date{\today}%


\begin{abstract}
Despite having outstanding electrical properties, graphene is unsuitable for optical devices because of its zero band gap. Here, we report two-dimensional excitonic photoluminescence (PL) from graphene grown on Cu(111) surface, which shows an unexpected remarkably sharp and strong emission near 3.16 eV (full-width at half-maximum $\leq$ 3meV) and multiple emissions around 3.18 eV. As temperature increases, these emissions blue-shift, showing the characteristic negative thermal coefficient of graphene. Observed PLs originate from significantly suppressed dispersion of excited electrons in graphene caused by hybridization of graphene $\pi$ and Cu \textit{d} orbitals of the 1st and 2nd Cu layers at a shifted saddle point 0.525(M+K) of Brillouin zone. This finding provides a new pathway to engineering novel optoelectronic graphene devices, whilst maintaining the outstanding electrical properties of graphene.
\end{abstract}

\maketitle

Graphene has been widely studied due to its remarkable electronic properties ~\cite{Liu2011:a,Novoselov2012:a,Grigorenko2012:a,Abajo2013:a,Vakil2011:a,Geim2004:a}. Nevertheless, owing to zero band gap, graphene has not been considered as useful optical materials \cite{Geim2004:a,Novoselov2004:a,Zhang2005:a,Novoselov2005:a,Geim2007:a,Mooradian1969:a}. Excited electron and hole pairs are easily screened by free electrons in metals. This makes the luminescence efficiency of metals very low so that only the band to band transition may be signified, as observed in noble metals \cite{Mooradian1969:a}. Surprisingly, excitonic features for metallic carbon nanotubes were predicted theoretically and observed in optical absorption experiments \cite{Spataru2004:a,Wang2007:a}, leading to possible enhancements in luminescence. These were explained by the fact that screening is not effective in one-dimensional metals. For 2-dimensional (2D) materials having intriguing 2D electronic features near the Fermi level like graphene, sharp luminescence was not obtainable. While an exciton in 2D semi-metallic graphene was predicted \cite{Yang2009:a,Yang2011:a}, no direct experimental evidence has been reported despite the existence of some signatures \cite{Mak2011:a}. Here we show the photoluminescence (PL) of graphene on Cu, which reveals the presence of an exciton in the quasi 2D system, where Cu is unique in the sense that it interacts weakly with graphene so that the Dirac cone remains.

\begin{figure*}
\centering
\includegraphics[width=17.2cm]{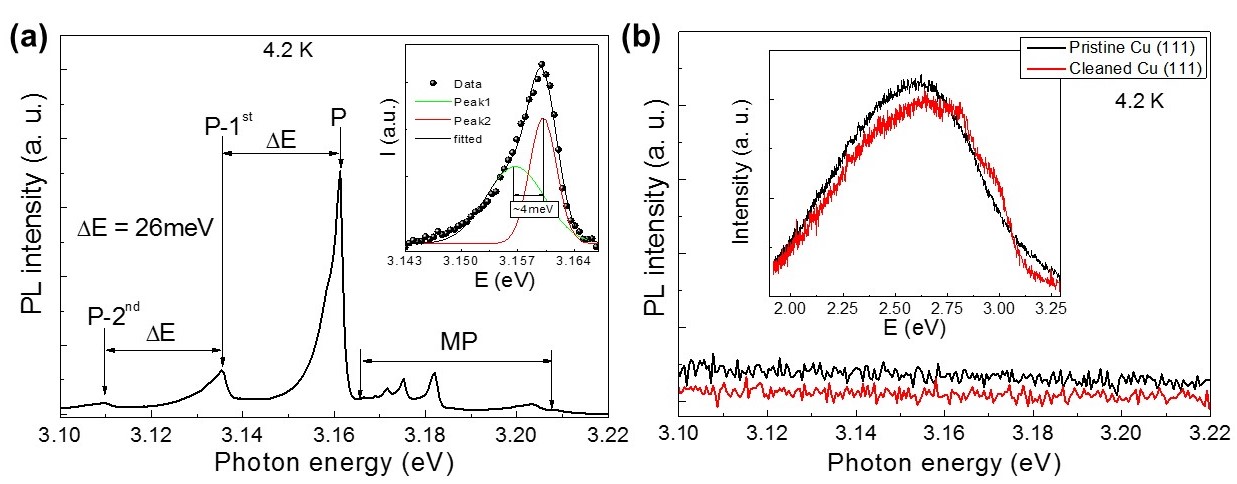}
\caption{\label{FIG. 1} (a) Micro-PL spectrum of graphene grown on a Cu (111) surface measured at an arbitrary position at 4.2 K. Inset depicts the reconstructed PL spectrum for the P emission using two Gaussian functions. (b) Micro-PL spectra of pristine and cleaned Cu(111) by annealing the surface in the CVD chamber prior to grow the graphene. Inset shows the wide micro-PL spectra of these samples.}
\end{figure*}

Graphene was grown on Cu single-crystal $7\times7\times1$ mm$^3$ in a hot furnace consisting of a 25 mm ID quartz tube. The Cu single-crystal disc was first placed in the center of a horizontal quartz tube mounted inside a high temperature furnace, and the tube was then evacuated, back filled with hydrogen (H$_2$) and argon (Ar), and the process was repeated three times to remove the residual air completely from the quartz tube. H$_2$ (100 sccm) and Ar (200 sccm) were introduced as the carrier gas, when the furnace temperature reached 1323 K. A Cu single-crystal was pre-annealed at 1323 K for 30 mins to remove the native Cu oxide layer in the H$_2$ and Ar atmosphere. Then, CH$_4$ (5 sccm) was flowed through the system. In order to know the relation of the PL with graphene, we also prepared pristine Cu(111) and cleaned Cu(111) obtained after annealing without introducing the CH$_4$ gas. The growth pressure was set to 5 Torr during growth where monolayer graphene is formed. After exposure to CH$_4$, the furnace was cooled to room temperature with H$_2$ and Ar.

The core-level photoemission spectroscopy studies were performed at a pressure of $1.0 \times 10^{ - 10}$ Torr at the 10D and 4A2 beamline of the Pohang Accelerator Laboratory (PAL), equipped with a PHOIBOS 150 electron energy analyzer with a 2D charge-coupled device (CCD) detector (Specs GmbH). All the spectra were collected at the normal emission. The photon energy used 360 eV to obtain high-quality C 1s core level spectra (Fig. S1). The binding energy scale was calibrated with the Au 4f core-level peak at 84.0 eV \cite{Moulder1995:a}.

For the plane wave DFT calculations, we used the local density approximation (LDA) exchange functional and a plane wave basis set with 600 eV cut-off energy. We sampled the BZ with a ($128\times128\times1$) k-point mesh to calculate the projected density of states. We used DFPT to calculate the phonon dispersion and the phonon DOS. We used the lattice constant of a pristine graphene for the combined system of graphene and Cu to simplify our calculation. The electronic coupling between the graphene and Cu(111) is weak based on the ARPES experiments \cite{Dedkov2001:a,Herrero2016:a,Dedkov2015:a} showing intact Dirac bands and from the theoretical calculation \cite{Vita2014:a,Voloshina2014:a} predicting the mean interlayer distance ~3.0 $\AA$ and the corrugation ~0.2 $\AA$. Thus, we used a weakly interacting distance 3 $\AA$ and a ($1\times 1$) unit cell after verifying insignificant effects of stacking on the electronic band structure (Fig. S2). The independent particle approximation (IPA) was used to compute the dipole transition matrix.

A frequency-tripled femtosecond Ti:sapphire laser (100 fs pulses at 76 MHz) operating at 266 nm was used to excite the graphene in the $\mu$-PL experiments. The sample was mounted in a continuous-flow helium cryostat, allowing the temperature to be controlled accurately from 4.2 K to room temperature. A 36x reflecting objective was held by a sub-micron precision piezoelectric stage above the cryostat and used to focus the incident laser beam to a spot size of ~2 $\mu m^2$ and to collect the resulting luminescence. The luminescence was then directed to a spectrometer with a spectral resolution of ~700 $\mu$eV. The signal was finally detected using a cooled charge coupled device (CCD) detector. All the PL spectra were obtained by using 1200 gr/mm of monochrometer, except for 300 gr/mm was used for taking wide range PL from the pristine and cleaned Cu(111).

\begin{figure}
\centering
\includegraphics[width=8.6cm]{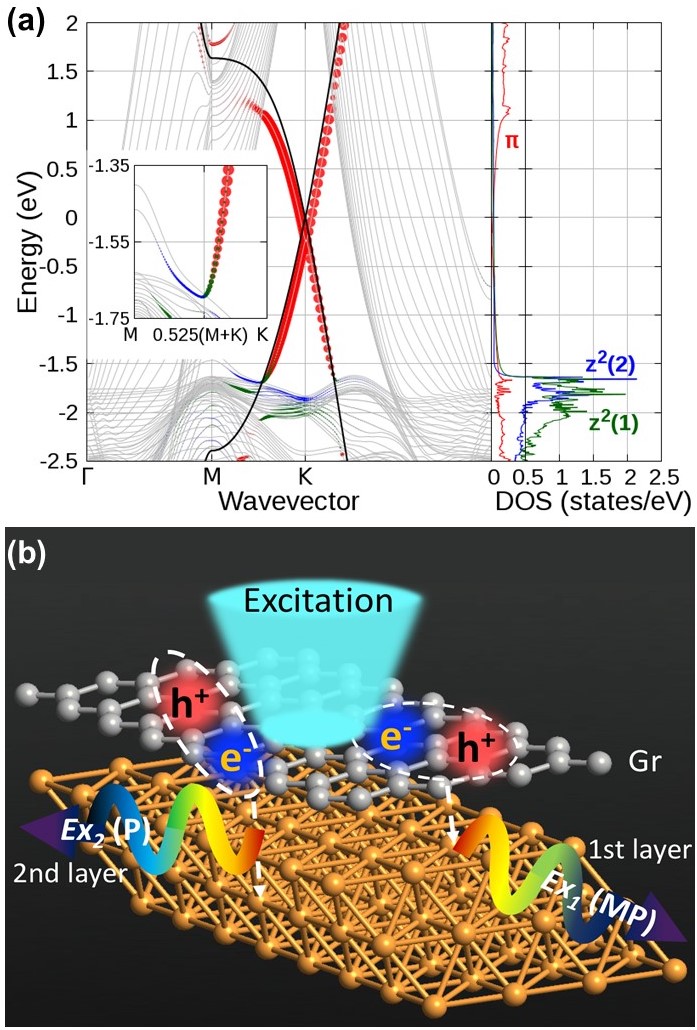}
\caption{\label{FIG. 2} (a) Band structure and projected DOS for 21-layer Cu (111)-graphene. Graphene $\pi$ (red), first layer Cu(1) \textit{d$_{z^{2}}$} (green), second layer Cu(2) \textit{d$_{z^{2}}$} (blue) and pristine graphene (black). Enlarged band structure around the vHs at 0.525 (M$+$K) in the valence band (Inset). (b) Schematic of the exciton formation and PL phenomena.}
\end{figure}

$\mu$-PL spectra of a graphene sheet synthesized on a Cu(111) surface measured at an arbitrary position is shown in Fig. 1(a). A strong and sharp PL peak near 3.161 eV (the lowest energy emission peak, denoted as P) and many peaks around 3.18 eV (denoted as MP) were observed. Fig. 1(b) shows that no PL is visible in the range of 3.1 $\sim$ 3.22 eV for a pristine Cu(111) and a cleaned Cu(111) in the CVD chamber, indicating that the PLs in Fig. 1(a) come from the graphene grown on the substrate. Unlike the PL of graphene quantum dots (GQDs) and graphene oxide (GO), which exhibit a broad luminescence \cite{Liu2013:a,Pan2010:a,Kim2012:a}, very sharp PL emission features showing a full-width at half-maximum (FWHM) $\Gamma$ of $\leq$ 3 meV were measured. These emissions have not been observed in ordinary graphene or GQDs. The asymmetric P peak can be resolved into two components by fitting with Gaussian functions. The two emission lines are reconstructed as shown in the inset of Fig. 1(a). The lower energy peak may originate from a phonon replica of the higher energy peak with a phonon energy shift of $\sim$4 meV (discussed later). We can rule out the possibility that the emission originates from graphene oxide or impurities. The reasons are: (i) the Raman G peak $\sim$1582 cm$^{-1}$ observed in the graphene on Cu(111) indicates a negligible doping level \cite{Yan2007:a} and there is no Raman peak related to graphene oxide in Figure S3(a), and (ii) the C 1s core level peak is very sharp with a FWHM of $\sim$0.8 eV near 284.5 eV with no noticeable oxide-related peaks, such as graphene oxide, COOH, CO$_2$, and CuO in the C 1s and Cu 3p core level spectra in Figure S1 for C 1s. Water molecules physisorbed on the surface in air seem to be mainly responsible for the O 1s peak observed in wider spectra taken with 630 eV. The experimental results are verified by density functional theory (DFT) and density functional perturbation theory (DFPT) calculations (discussed later). Apart from the above emissions, additional peaks near 3.135 eV (denoted as P-1st) and 3.109 eV (denoted as P-2nd) were observed. These peaks are identified as 1st and 2nd phonon replica of P with a phonon energy of $\sim$26 meV (as discussed in Figs. 3 and 4).

To understand the physics of these peaks, we performed DFT calculations using the VASP suite \cite{Kresse1996:a} with local density approximations. We consider three possible stacking configurations, where graphene carbon atoms in a unitcell (A and B) sit on Cu atoms of 1st and 2nd layers (tophcp), 1st and 3rd layers (topfcc), or 2nd and 3rd layers (hcpfcc). Geometry optimization for different stackings showed weak interaction of graphene with Cu, d $>$ 3.0 $\AA$, in agreement with the very weak interaction based on the accurate ab initio calculations \cite{Youn2012:a}. The overall feature of band structure remains consistent in various stacking configurations (Fig. S2). Here, we use one of stacking configurations, tophcp. The Dirac band (Fig. 2) of graphene (red dots) remains almost intact with little downshift as compared to the pristine case (black line). Copper bands contain dispersive \textit{sp} conduction band and flat \textit{d} valence band (blue and green dots). Although graphene interacts with copper weakly, the graphene $\pi$ band is significantly perturbed by Cu \textit{p$_x$} near the M point of Brillouin zone (BZ). This produces a new van Hove singularity (vHs) with the flattened band near the mid-point of M-K at an energy of 1.1 eV, giving rise to an enhanced density of states (DOS) as shown in DOS plot on the right side of Fig. 2. For the valence band, the top-most Cu(1) \textit{d$_{z^{2}}$} band (near -1.6 eV) shows a significant orbital hybridization (see the inset) with the $\pi$ band of graphene, consistent with the literature \cite{Vita2014:a}, and produces large DOS peaks. Such an enhanced DOS for electrons and holes is likely to support an emissive transition due to (i) a significant coupling between C $\pi$ and Cu \textit{d$_{z^{2}}$} orbitals, (ii) an optical selection rule and (iii) an insignificant change in momentum (the direct transition). We calculated the transition probability $P(\omega) \propto A_{fi} \rho_f \rho_i$, where $A_{fi}$ is the dipole transition matrix between initial and final states and $\rho_{i(f)}$ is the density of initial (final) state. The DFT result shows a peak with a transition energy of 2.7 eV, which corresponds to the expected emission energy of P (Fig. S4). Small discrepancies of DFT-predicted energy gaps are generally expected, as the energy gap of bulk Si is underestimated by 0.61 eV \cite{Perdew1985:a}. 

\begin{figure}
\centering
\includegraphics[width=8.6cm]{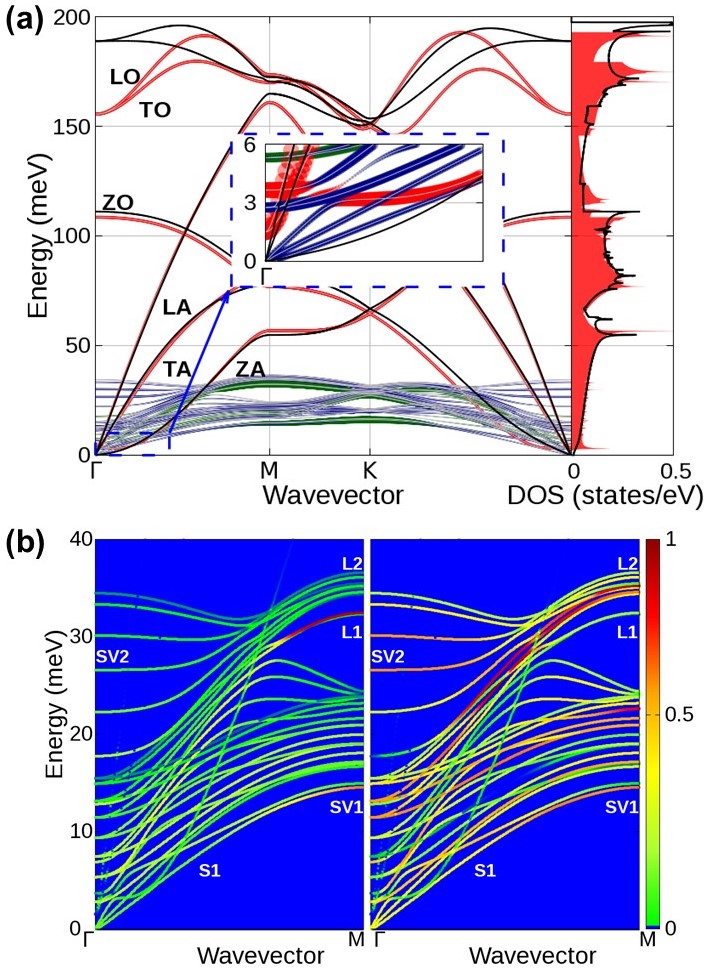}
\caption{\label{FIG. 3} (a) Dynamical matrix projected phonon dispersion and phonon DOS of 9-layer Cu (111)-graphene, pristine graphene (black), graphene (red), the first Cu layer (green) and the second layer Cu (blue). (b) Normalized dynamical matrix projected phonon dispersion of the first (left) and second (right) layers near $\Gamma$.}
\end{figure}

The possibility that lattice vibrations are responsible for the asymmetric P line and its replica was investigated using PHONOPY package \cite{Togo2012:a} for the phonon dispersions and DOS of graphene on 9 Cu(111) layers. As shown in Fig. 3(a), the DOS has a peak at 3.65 meV due to the suppressed dispersion of the out-of plane acoustic phonon (ZA) mode (see the inset) near the $\Gamma$ point of the BZ. It describes an asymmetric P composed of consecutive overlapping Gaussian peaks shifted by 3.65 meV. Coupling of the conduction band with the graphene ZA mode in Fig. S5(a) demonstrates the origin of the linear increase of the FWHM with the temperature, $\Gamma(T) \propto T$. The P replicas arise from the Cu surface phonon modes. The contributions of the first layer Cu(1) and the second layer Cu(2) to the low energy modes are shown in Fig. 3(b). The Cu(2) out-of-plane vibration mode SV2 \cite{Benedek2010:a} has an energy of 30 meV near the Γ point. Upon modulation of the SV2 mode, the electronic energy level Cu(2) \textit{d$_{z^{2}}$} increases by a significant amount in Fig. S5(b). This indicates a large electron-phonon coupling between the SV2 and Cu(2) \textit{d$_{z^{2}}$}. Our calculated phonon energy of 30 meV is comparable to the energy shift of 26 meV for the P replica. Therefore, P-1st and P-2nd correspond to Cu surface phonon emissions.

\begin{figure}
\centering
\includegraphics[width=8.6cm]{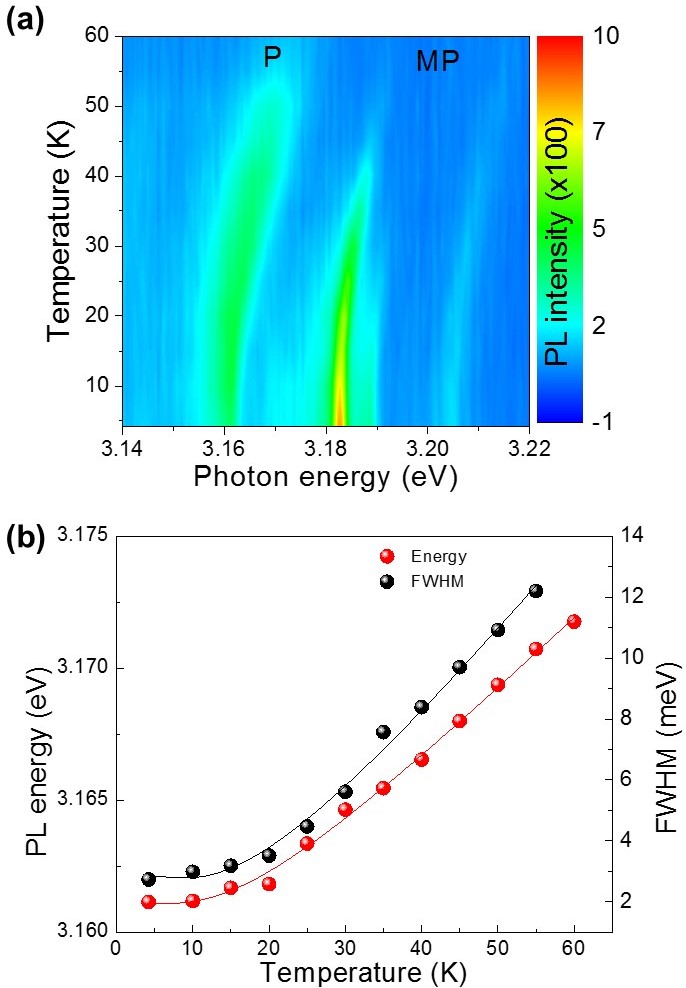}
\caption{\label{FIG. 4} (a) Temperature-dependent micro-photoluminescence spectra of the graphene measured at an arbitrary region. (b) PL peak energy and full width at half maximum for the P peak as a function of temperature. The solid lines are the least square fitted curves.}
\end{figure}

The PL spectra shift to higher energy with increasing temperature (Fig. 4). Similar emission peaks were observed for graphene grown on Cu foils, even in the case of low quality graphene (Fig. S6). The integrated intensity of the MP emissions decreases linearly with temperature, while that of the P emission increases up to $\sim$40 K and then decreases. Figure 4(b) shows both PL energy and FWHM for P and MP emissions as a function of temperature. This temperature dependent band gap variation is understood in terms of lattice dilation and electron-lattice interactions. Interestingly, the PL peaks shift to higher energy with increasing temperature, which is the opposite trend to other semiconducting materials. This is in agreement with the previous study verifying a negative thermal coefficient for graphene \cite{Yoon2011:a,Jiang2009:a,Zakharchenko2009:a}. The temperature dependence of band-gap proposed by O’Donnell and Chen \cite{Donnell1991:a} takes into account the influence of phonons on the bandgap energy to obtain a better fit for semiconductors at lower temperatures. They conisdered the following equation: $E_g(T) = E_g(0) - S <E_{ph}>[coth(<E_{ph}>/2k_{B}T)-1]$, where $<E_{ph}>$ is an average phonon energy and S is a dimensionless coupling constant. The measured data are in good fit with the aforementioned relationship at all measured temperatures (red solid lines). The fitting parameter of the $<E_{ph}>$ was found to be $\sim$26 meV for the P emission.

The FWHM of the P emission exhibits a small linear increase for temperatures up to $\sim$20 K, and then broadens with further increase in temperature. The thermal broadening of the emission line width due to the exciton-phonon interaction can be expressed by the equation of $\Gamma(T) = \Gamma_0 + \sigma T + \gamma_{LO} exp (-E_{LO}/k_BT)$ \cite{Seguin2004:a}, which is denoted as a black solid line in Fig. 4(b). Here $\Gamma_0$ is the temperature-independent inhomogeneous broadening; as T $\to$ 0 K, $\Gamma_0$ is 2.9 meV for the P emission from the fit. The last two terms are related to the homogenous broadening due to exciton-phonon interactions. Here $\sigma$ is the coupling coefficient between an exciton and an acoustic phonon, and $\gamma_{LO}$ is the coefficient for coupling between an exciton and the longitudinal optical (LO) phonon. The best fitting parameter is $\sigma$ = 313.6 $\mu$eV/K, much higher than the values reported usually for semiconductor materials with similar transition energies such as InGaN quantum dot (QD) (1.7 $\mu$eV/K) \cite{Seguin2004:a}, CdSe nanosheets (9.8 $\mu$eV/K for 4 ML) \cite{Achtstein2012:a}, and GaN QD (0.8 $\mu$eV/K) \cite{Amloy2011:a}, which means that the out-of plane acoustic phonon dominates over optical phonons. 

This work reports photoluminescence from a graphene sheet grown on a Cu surface. Strong and sharp emission lines were clearly observed near 3.16 eV (P) and 3.18 eV (MP). These emissions shift to higher energy with increasing temperature, indicating a negative thermal coefficient for graphene. From DFT calculations, the orbital hybridization of graphene and the Cu surface is responsible for the large optical transition probability and the shifting of the saddle point of the BZ. The present results could be utilized for development of new optoelectronic devices.

\begin{acknowledgments}
This research was supported by Basic Science Research Program (2015R1D1A1A01058332), the SRC Center for Topological Matter (No. 2011-0030787), and National Honor Scientist Program (2010-0020414) through the National Research Foundation of Korea (NRF). The experiments at PLS were supported in part by MSIP and POSTECH. Computation was supported by KISTI (KSC-2015-C3-059, KSC-2015-C3-061).
\end{acknowledgments}

\footnotetext[1]{contributed equally to this work.}

\bibliographystyle{apsrev4-1}
\bibliography{parkbib}

\end{document}